\documentclass[reprint,amsmath,amssymb,aps,preprintnumbers,nofootinbib,showpacs]{revtex4-1}

\usepackage{amssymb}
\usepackage{amsmath}
\usepackage{nicefrac}
\usepackage{graphicx}
\usepackage{subfigure}
\usepackage{fullpage}
\usepackage{colortbl}
\usepackage{hyperref}

\newcommand{\none}{$\varnothing$}

\newcommand{\resp}{\!\!\parbox{5mm}{\begin{fmffile}{resplotp}
      \setlength{\unitlength}{1mm}
      \begin{fmfgraph*}(3.0,3.0)
        \fmfleft{i1}
        \fmfright{o1,o2}
        \fmf{plain}{i1,v1}
         \fmfset{dot_len}{3}
       \fmf{dbl_dots}{v1,o1}
        \fmf{dbl_dots}{v1,o2}
      \end{fmfgraph*}
    \end{fmffile}}\!\!}
    
 \newcommand{\resone}{\!\!\parbox{5mm}{\begin{fmffile}{resplot1}
      \setlength{\unitlength}{1mm}
      \begin{fmfgraph*}(3.0,3.0)
        \fmfleft{i1,i2}
        \fmfright{o}
        \fmf{plain}{i1,v1}
       \fmf{plain}{i2,v1}
        \fmfset{dot_len}{3}
        \fmf{dbl_dots}{v1,o}
      \end{fmfgraph*}
    \end{fmffile}}\!\!}

 \newcommand{\restwo}{\!\!\parbox{5mm}{\begin{fmffile}{resplot2}
      \setlength{\unitlength}{1mm}
      \begin{fmfgraph*}(3.0,3.0)
        \fmfleft{i1,i2}
        \fmfright{o1,o2}
        \fmf{plain,tension=1}{i1,v1,v2,i2}
       \fmf{plain}{v1,o1}
       \fmf{plain}{v2,o2}
       \fmffreeze
       \fmfset{dot_len}{3}
         \fmf{dbl_dots}{v2,o2}
      \end{fmfgraph*}
    \end{fmffile}}\!\!}
    
     \newcommand{\resthree}{\!\!\parbox{5mm}{\begin{fmffile}{resplot3}
      \setlength{\unitlength}{1mm}
      \begin{fmfgraph*}(3.0,3.0)
        \fmfleft{i1,i2}
        \fmfright{o1,o2,o3}
        \fmf{plain,tension=2}{i1,v1,v2,v3,i2}
       \fmf{plain}{v1,o1}
       \fmf{plain}{v3,o3}
       \fmfset{dot_len}{3}
         \fmf{dbl_dots}{v2,o2}
      \end{fmfgraph*}
    \end{fmffile}}\!\!}

      \newcommand{\resthreea}{\!\!\parbox{5mm}{\begin{fmffile}{resplot3a}
      \setlength{\unitlength}{1mm}
      \begin{fmfgraph*}(3.0,3.0)
        \fmfleft{i1,i2}
        \fmfright{o1,o2,o3}
        \fmf{plain,tension=2}{i1,v1,v2,v3,i2}
       \fmf{plain}{v1,o1}
       \fmf{plain}{v2,o2}
       \fmfset{dot_len}{3}
         \fmf{dbl_dots}{v3,o3}
      \end{fmfgraph*}
    \end{fmffile}}\!\!}
      
       \newcommand{\resfour}{\!\!\parbox{5mm}{\begin{fmffile}{resplot4}
      \setlength{\unitlength}{1mm}
      \begin{fmfgraph*}(3.0,3.0)
        \fmfleft{i1,i2}
        \fmfright{o1,o2,o3,o4}
        \fmf{plain,tension=2}{i1,v1,v2,v3,v4,i2}
       \fmf{plain}{v1,o1}
       \fmf{plain}{v4,o4}
       \fmfset{dot_len}{3}
         \fmf{dbl_dots}{v3,o3}
             \fmffreeze
       \fmf{plain}{v2,o2}
      \end{fmfgraph*}
    \end{fmffile}}\!\!}
       
        \newcommand{\resfive}{\!\!\parbox{5mm}{ \begin{fmffile}{resplot5}
      \setlength{\unitlength}{1mm}
      \begin{fmfgraph*}(3.0,3.0)
        \fmfleft{i1,i2}
        \fmfright{o1,o2,o3,o4,o5}
        \fmf{plain,tension=2}{i1,v1,v2,v3,v4,v5,i2}
       \fmf{plain}{v1,o1}
       \fmf{plain}{v5,o5}
       \fmfset{dot_len}{3}
         \fmf{dbl_dots}{v3,o3}
       \fmffreeze
       \fmf{plain}{v2,o2}
          \fmf{plain}{v4,o4}
      \end{fmfgraph*}
    \end{fmffile}}\!\!}

\newcommand{\bnum}[2]{( \mathbf{#1},#2)}
\newcommand{\fnum}[2]{[\mathbf{#1},#2]}

\usepackage{feynmp-auto}
\begin{document}

\preprint{UCI-TR-2015-XX}
\preprint{PITT-PACC-1610}

\title{The unexplored landscape of two-body resonances}

\begin{abstract}
We propose a strategy for searching for theoretically-unanticipated new physics which avoids a large trials factor by focusing on experimental strengths.  Searches for resonances decaying into pairs of visible particles are experimentally very powerful due to the localized mass peaks and have a rich history of discovery. Yet, due to a focus on subsets of theoretically-motivated models, the landscape of such resonances is far from thoroughly explored.  We survey the existing set of searches, identify untapped experimental opportunities and discuss the theoretical constraints on models which would generate such resonances.
\end{abstract}

\author{Nathaniel Craig}
\affiliation{Department of Physics, UC Santa Barbara, Santa Barbara, CA 93106}
\author{Patrick Draper}
\affiliation{Department of Physics, University of Massachusetts, Amherst, MA 01003}
\author{Kyoungchul Kong}
\affiliation{Department of Physics and Astronomy, University of Kansas, Lawrence, KS 66045}
\affiliation{Pittsburgh Particle physics, Astrophysics, and Cosmology Center, Department of Physics and
Astronomy, University of Pittsburgh, Pittsburgh, PA 15260}
\author{Yvonne Ng}
\affiliation{Department of Physics and Astronomy, UC Irvine, Irvine, CA 92627}
\author{Daniel Whiteson}
\affiliation{Department of Physics and Astronomy, UC Irvine, Irvine, CA 92627}

\date{\today}

\maketitle

\section{Introduction}

Searches for two-body decays of heavy resonances have a rich history of important discoveries, from the $J/\psi$ to the Higgs boson.  Such resonances can provide an unambiguous signature of a localized invariant mass peak and offer simple background estimation from sidebands, allowing for discovery without requiring full models of the signal or background processes.  These experimental features, combined with compelling theoretical arguments, motivate much of the current program of resonance searches.

The theoretical arguments for new resonances mostly consist of simple generic extensions to the Standard Model (e.g. a new $U(1)$) or modifications to the SM which address an outstanding theoretical problem (e.g. Kaluza-Klein gravitons).  To date, most of the experimental searches have followed these theoretical arguments, leading to many searches for pairs of identical objects (eg $ee,\mu\mu, jj$) and in rarer cases for non-identical pairs (eg $e\mu,ZW$).   However,  the dramatic scale of the open theoretical  questions facing particle physics suggests that a correct theory of Nature may not be one of the models currently in fashion or under specific consideration.   This motivates an experimental program which is not narrowly focused on current models and the signatures they suggest, but with a broad scope and systematic approach capable of theoretically unanticipated discoveries. While there have been many proposals for model-independent search programs at hadron colliders (such as the framework of on-shell effective theories \cite{ArkaniHamed:2007fw}), they have been largely motivated by specific theoretical frameworks, and consequently many holes remain in the existing experimental program at the LHC. To make concrete progress, we propose a systematic search for new particles decaying into $n$-body resonances. In the $n=2$ case, this would consist of searches for resonances in all pairs of objects, even those which have no theoretical motivation or are theoretically disfavored.

The typical difficulty facing searches without specific theoretical motivation is the large number of possible observables, which incurs a very large trials factor and greatly reduces the discovery sensitivity.  Here, rather than relying on theoretical guidance, we propose to restrict the vast space of possible theories into those that align well with experimental strengths.    We are interested in covering the intermediate ground between the very specific and the very general search programs, by focusing on well-defined topologies independent of specific theory considerations. This broadens the search program beyond favored theories, but not so much so as to compromise discovery potential.  Given that the data exist and resonances are fairly easy to discover, we argue that the two-particle spectra are worth directly examining. In many cases, there are indirect constraints on such resonances from other experiments or subjective theoretical arguments, but there is no real substitute for a direct search.

In this paper, we lay out the details of the implementation of such a search program and survey the existing experimental and theoretical landscape for exclusive $n=2$-body resonances, leaving $n=3+$ (as well as inclusive $n=2$ final states) for future work.  We find that the majority of 2-body resonances have some indirect theoretical constraints but have received almost no experimental attention, leaving most of the landscape unexplored and a large potential for unanticipated discovery. 

\section{Scope \& experimental searches}

We consider resonances decaying to a basic set of identifiable light objects (charged leptons, photons, light-quark jets, $b$-tagged jets) as well as heavy objects (top quarks, weak bosons, Higgs bosons) which are routinely identified\footnote{One could imagine restricting the scope to light objects, categorizing the heavy objects as higher-level decays (eg $X\rightarrow WW\rightarrow 4j$ would be considered in the $n=4$ category rather than $X\rightarrow WW$ as $n=2$). This is equivalent, but allows us to call attention to these typical objects rather than considering them as special mass cases of higher-level decays.}. In the case of $n=2$ objects, this gives 55 unique pairs of exclusive final states, see Table~\ref{tab:res}.  Final states with higher number of objects have a larger number of exclusive final states; we reserve these for future work.

We examined experimental searches from ATLAS and CMS in data collected from proton-proton collisions with $\sqrt{s}=8$~TeV.  We consider exclusive final states only in terms of the pairs of identifiable objects defined above. For example, in the $e\gamma$ category of this exclusive $n=2$ survey, we consider only searches for $e\gamma$, of which there are none, and do not consider searches for $e^+e^-\gamma$, of which there are several motivated by excited lepton models that give a resonance in $e\gamma$. The final state of $e^+e^-\gamma$ would be covered by an $n=3$ study, and extrapolation of those limits to the $n=2$ $e\gamma$ category requires theoretical assumptions about the production modes.

The survey of $n=2$ final states is shown in Table~\ref{tab:res}, with the striking feature that most diagonal entries have existing searches, where as most off-diagonal entries do not. In the case of the Higgs boson in particular, there are several unexamined resonance categories. Note that the lack of searches in these resonance categories is not for want of theory models. Examples of theories that populate the entire landscape of 2-body resonances are shown in Table~\ref{tab:restheory}.

\begin{figure*}
\includegraphics[width=0.7\textwidth]{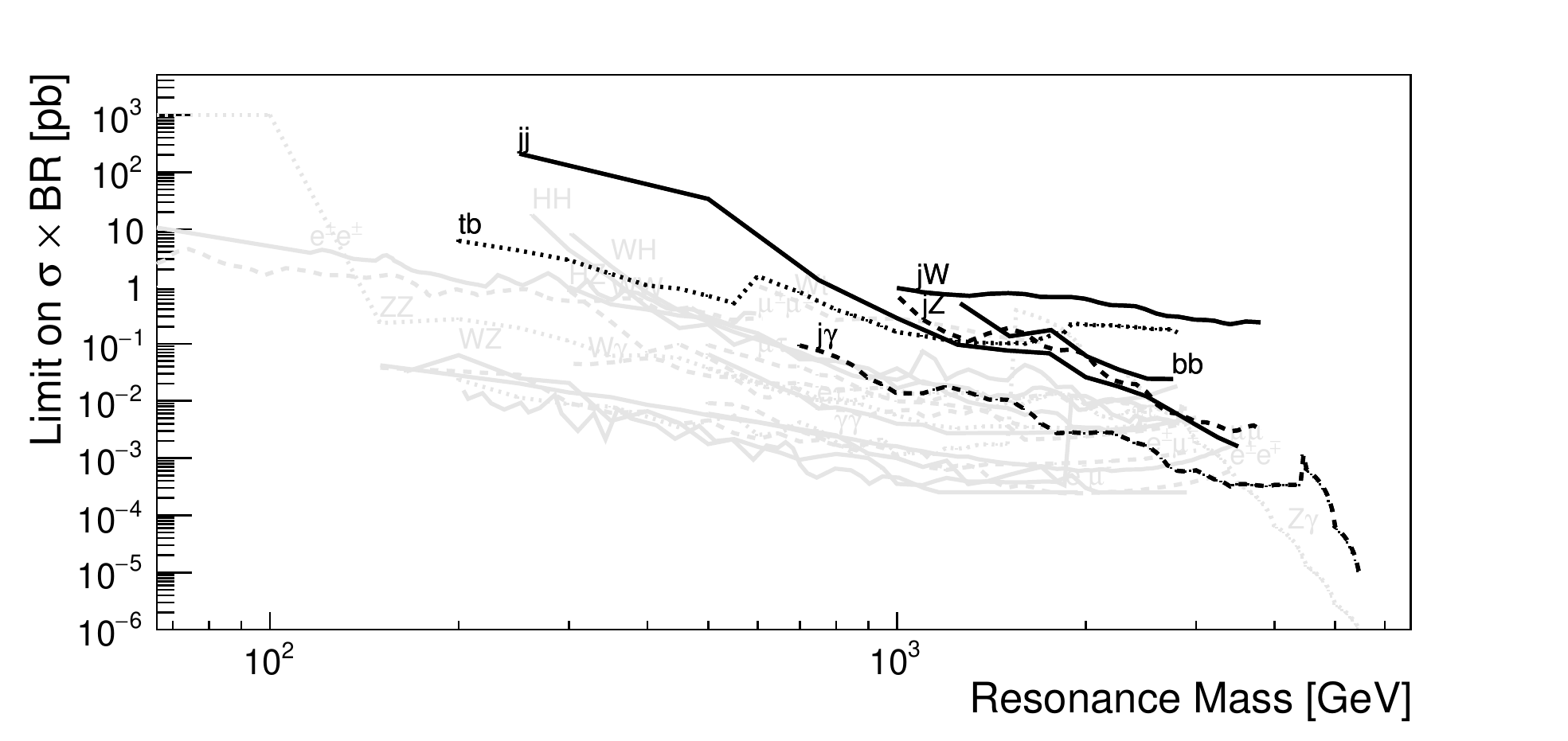}
\includegraphics[width=0.7\textwidth]{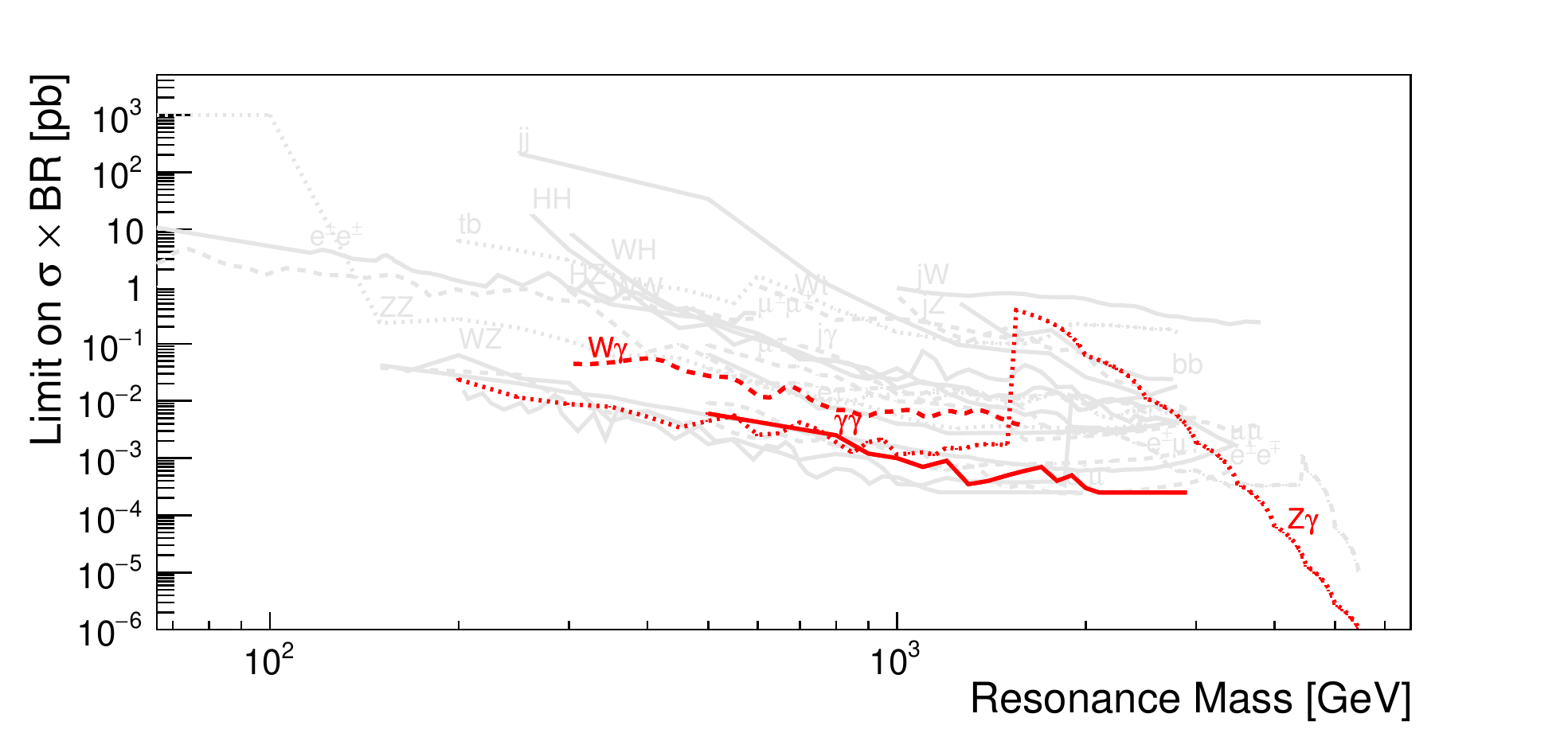}
\caption{ Existing limits on the cross section times branching ratio for resonances to various 2-body final states, as a function of the resonance mass. Top pane emphasizes hadronic final states, bottom pane emphasizes photonic final states. References for searches can be found in Table~\ref{tab:res}.}
\label{fig:jg}
\end{figure*}

\begin{figure*}
\includegraphics[width=0.7\textwidth]{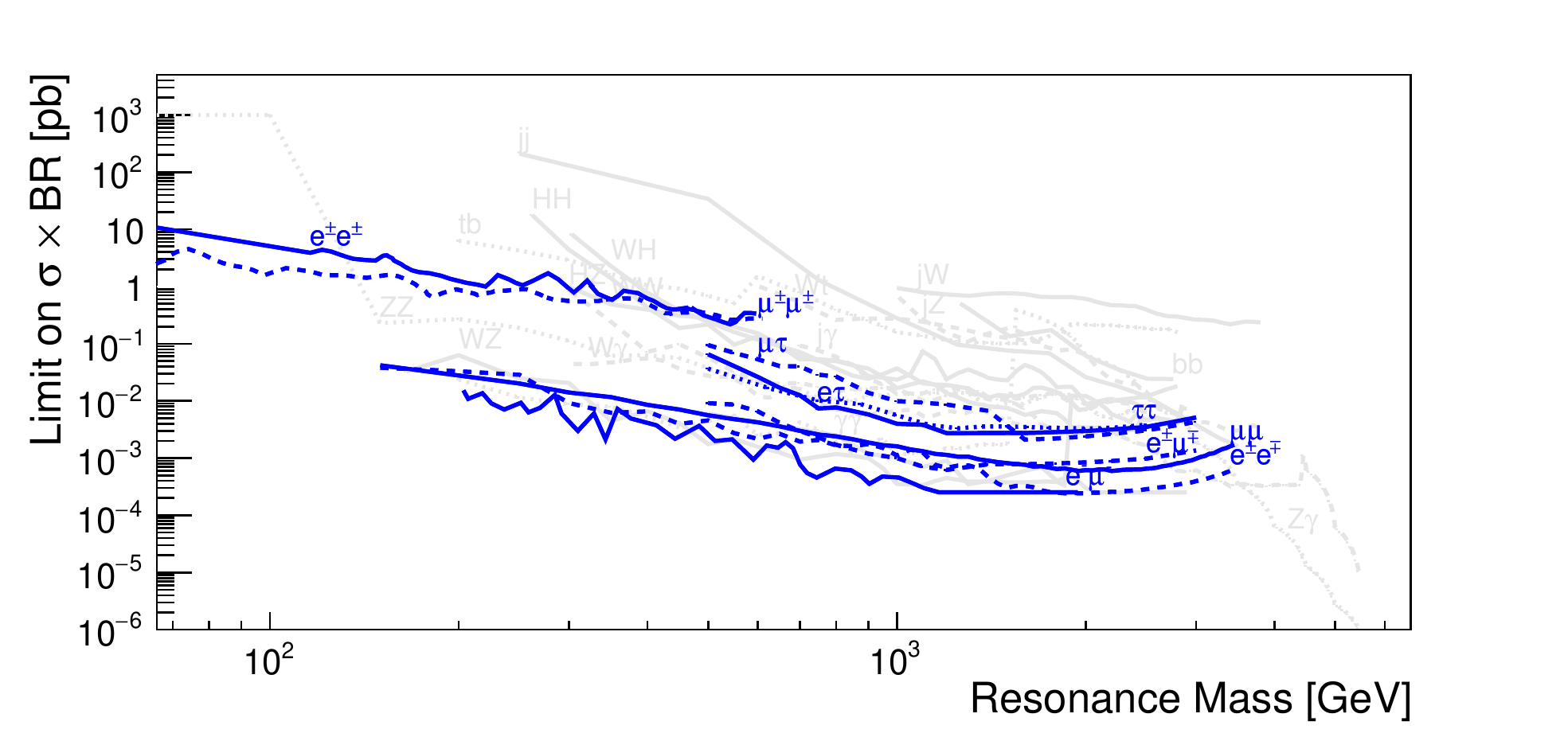}
\includegraphics[width=0.7\textwidth]{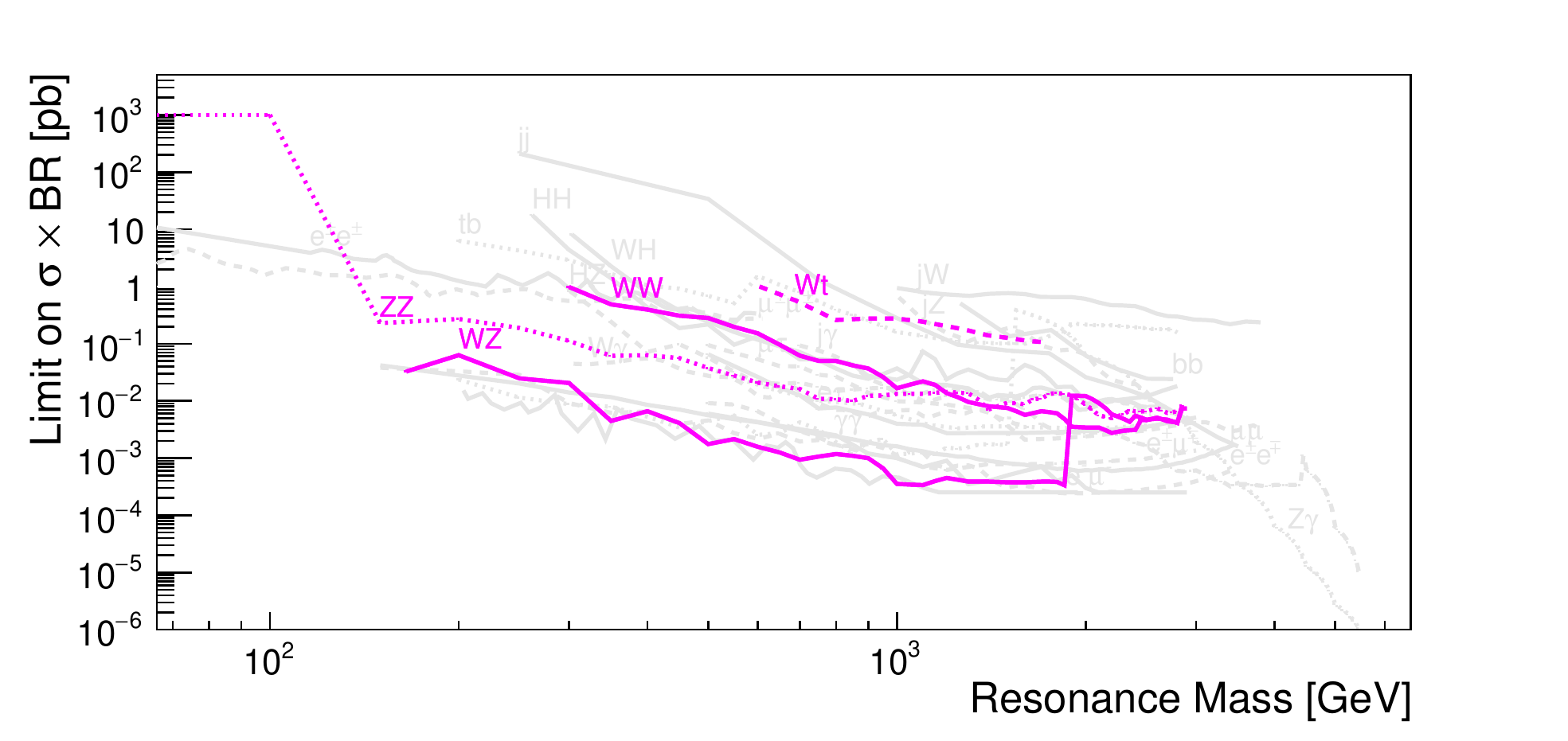}
\includegraphics[width=0.7\textwidth]{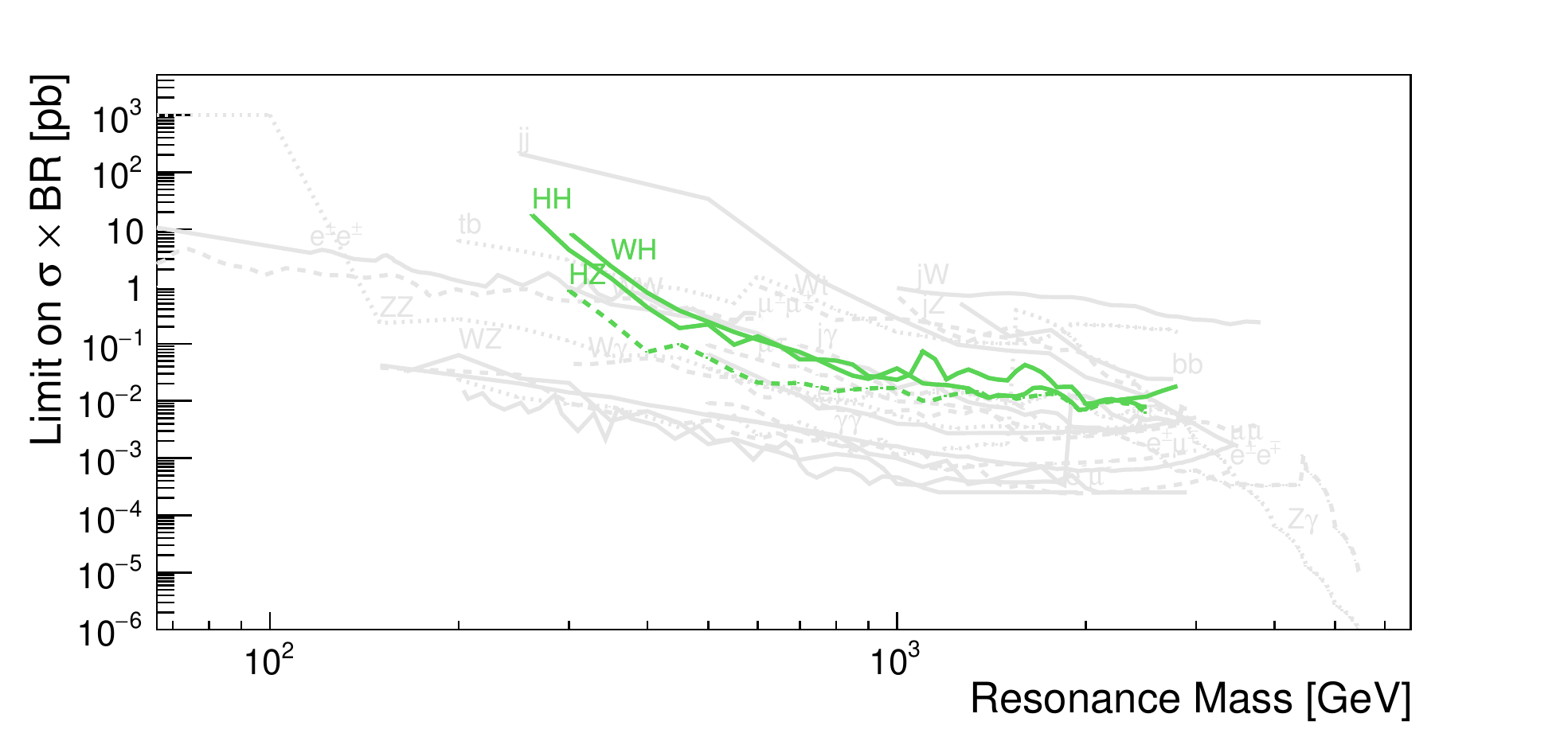}
\caption{ Existing limits on the cross section times branching ratio for resonances to various 2-body final states, as a function of the resonance mass. Top pane emphasizes leptonic final states, center pane emphasizes bosonic final states, and the bottom pane emphasizes Higgs final states. References for searches can be found in Table~\ref{tab:res}.}
\label{fig:lwh}
\end{figure*}

Even in cases where searches exist, there are often unexamined regions in the resonance mass. Figures~\ref{fig:jg} and~\ref{fig:lwh} show the strongest limits on the cross section times branching ratio as a function of the resonance mass for all results which satisfy the requirements.

\section{Theoretical constraints}

Various theoretical constraints may be imposed on $n$-body resonances, which in turn influence the likely production and decay modes at the LHC. In order to maintain the broadest possible scope, we consider only the most stringent constraints imposed by gauge invariance and Lorentz invariance, as experimental constraints on e.g.~flavor violation depend on the details of the underlying model and may in principle be evaded.

Gauge invariance and Lorentz invariance restrict the possible statistics and quantum numbers of a resonance decaying to a specified 2-body final state. The statistics and possible $SU(3)_c$ and $U(1)_{em}$ numbers of 2-body resonances are enumerated according to their exclusive final state in Table~\ref{tab:quantumnumbers}. Note that we enumerate only $SU(3)_c \times U(1)_{em}$ quantum numbers rather than $SU(3)_c \times SU(2)_L \times U(1)_Y$ quantum numbers, because a large number of $SU(3)_c \times SU(2)_L \times U(1)_Y$ representations may share the same exclusive final state provided additional insertions of the Higgs vacuum expectation value. We also do not exhaustively list all possible $SU(3)_c$ representations, but for simplicity restrict our attention to states transforming in the fundamental or adjoint representation; resonances transforming in other representations of $SU(3)_c$ may have different pair production cross sections but do not lead to significantly different signatures. While a fermionic resonance with Standard Model quantum numbers generally contributes to gauge anomalies, these anomalies may be cancelled by additional particles that do not influence the collider signatures of the resonance. 

Gauge invariance and Lorentz invariance also dictate the structure of operators coupling a resonance to Standard Model particles, and in many cases the couplings must arise via irrelevant operators. For example, a resonance $X$ decaying to $tg$ cannot couple via a minimal gauge coupling $ \bar X \gamma^\mu G_\mu t$, but may couple via e.g.~a chromoelectric dipole operator of the form $\bar X \gamma^{\mu \nu} G_{\mu \nu} t$. In many cases, more than one Lorentz structure is allowed for a given coupling. The various possible Lorentz structures for each coupling have a modest impact on kinematic distributions for the production and decay of each resonance (see e.g.~\cite{ArkaniHamed:2007fw}), but they do not alter the key feature of interest in this work, namely a bump in the $n$-body invariant mass spectrum. 

Note that these conclusions may be altered in the presence of significant interference effects, which may lead to deficits or peak-dip structures in the invariant mass spectrum if the Standard Model continuum interferes with the signal process. The existence and structure of interference effects cannot be determined by quantum numbers alone, and depends additionally on both the Lorentz structure and phases of couplings between the resonance and Standard Model states. However, in the limit of weak coupling, interference between a narrow resonance and Standard Model continuum backgrounds is negligible and may be neglected. To good approximation, as an expansion at weak coupling, searches for $n$-body resonances may therefore be parameterized solely in terms of the resonance mass, width, and production cross section times branching ratio.

Having specified the possible gauge quantum numbers of the 2-body resonance given the final state, gauge invariance and Lorentz invariance provide a guide to the possible production modes at the LHC. For each resonance there are three possibilities:

\begin{enumerate}
\item The particle can be {\it resonantly produced} either exclusively using its tree-level decay coupling (as in, e.g., a resonance decaying to $qq$ or $gg$); via loop-induced processes involving the decay coupling (as in, e.g., gluon fusion production of a $t \bar t$ resonance); or via additional couplings to quarks and gluons allowed by its quantum numbers. The presence of such additional couplings may lead to additional theoretical constraints discussed below. Such resonant production channels fall under the scope of the exclusive 2-body searches proposed here.
\item The particle can be produced via {\it associated production} exclusively using its decay couplings. For example, a resonance $X$ coupling to $tW^+$ can be produced in the process $q g \to t q X$ using only the $X t W^+$ coupling and Standard Model gauge couplings. This assumes no additional couplings to quarks and/or gluons. Such associated production channels fall under the scope of $n \geq 3$  studies, with a feature in the appropriate 2-body invariant mass spectrum.
\item The particle can be {\it pair produced} using its gauge quantum numbers (e.g. Drell-Yan via  electroweak quantum numbers). This process is kinematically suppressed for heavier resonances, but may be appreciable if the gauge couplings are significantly larger than the decay couplings. Such pair production channels fall under the scope of $n=4$ studies, with features in the appropriately-paired 2-body invariant mass spectra.
\end{enumerate}

The possible production modes for each resonance are enumerated in Table \ref{tab:modes}. In principle, a given resonance may be produced in all three modes, with varying rates depending on the relative sizes of phase space factors and production and decay couplings. In each case the final state contains a peak in the appropriate 2-body invariant mass, but with varying amounts of additional event activity. In this sense, the associated- and pair-production modes may not qualify for the $n=2$ exclusive case considered above, but serve as a useful foundation for future $n>2$ studies.

As is apparent in Table \ref{tab:modes}, there are several possible 2-body resonances for which resonant production is incompatible with Standard Model gauge invariance, in the sense that the quantum numbers of the final state cannot be produced by any initial state with appreciable parton density in proton-proton collisions. Nonetheless, searches for these 2-body resonances at the LHC remain motivated by the possibility of new physics that mimics a Standard Model final state in the LHC detectors (in the sense that, e.g., a long-lived neutral particle decaying to electron-positron pairs might be reconstructed as a photon). These states may also be produced in associated production with associated particles sufficiently soft to still appear as an exclusive 2-body resonance, or may originate from $n \geq 2$ exclusive final states with missing energy appearing in $n=2$ exclusive searches. Such states may also be resonantly produced at other colliders consistent with gauge invariance, such as in electron-proton collisions at HERA.

Apart from gauge invariance and Lorentz invariance, less robust constraints may also apply. Many such constraints arise only when the resonance possesses both its decay coupling and additional couplings to quarks and/or gluons. Proton decay provides the strongest such constraint, as strong bounds on the proton lifetime imply that the couplings of resonances inducing proton decay are vanishingly small. In the case of 2-body resonances, resonances coupling to a single pair of Standard Model particles will not induce proton decay, but proton decay may be induced by additional couplings to quarks required for resonant production at the LHC. Resonances for which this occurs are indicated in Table~\ref{tab:quantumnumbers}; in these cases it is reasonable to expect $n=2$ resonant production rates to be small.

Beyond proton decay, there are a variety of constraints on flavor violation, lepton number violation, and other types of baryon number violation, but in practice even strong constraints may be avoided by appropriate symmetries, textures, or fortuitous cancellations (as in e.g.~maximal flavor violation \cite{BarShalom:2007pw} or diquark-type interactions \cite{Giudice:2011ak}). In these cases there is no substitute for a direct search.

\begin{table*}
\caption{ Existing two-body exclusive final state resonance searches at $\sqrt{s}=8$ TeV.  The \none\ symbol indicates no existing search at the LHC.}
\begin{tabular}{lccccccccccc}
\hline
\hline
 & $e$\ \   & $\mu$\ \   & \ \  $\tau$\ \   & \ \  $\gamma$\ \   & \ \  $j$ \ \   & \ \  $b$\ \  & \ \ $t$ \ \ & \ \ $W$ \ \ & \ \ $Z$ \ \ & \ \ $h$ \ \   \\
\hline
$e$ & $\pm\mp$\cite{atlasdilepton8tev},$\pm\pm$\cite{atlassslep8tev} & $\pm\pm$\cite{atlassslep8tev,Khachatryan:2016ovq} $\pm\mp$\cite{Aad:2015pfa,Khachatryan:2016ovq}  & \cite{Aad:2015pfa} & \none  & \none & \none& \none& \none& \none& \none\\
$\mu$ & & $\pm\mp$\cite{atlasdilepton8tev},$\pm\pm$\cite{atlassslep8tev} &   \cite{Aad:2015pfa}& \none & \none & \none & \none& \none& \none& \none\\
$\tau$ & & & \cite{atlastautau8tev} & \none & \none & \none & \cite{Khachatryan:2015bsa}& \none& \none& \none\\
$\gamma$ & & & & \cite{atlasdiphoton8tev} & \cite{cmsphotonjet8tev,atlasphotonjet8tev,Aad:2015ywd} & \none & \none & \cite{Aad:2014fha} & \cite{Aad:2014fha} & \none\\
$j$ & & & & & \cite{atlasdijet8tev} & \cite{CMS-PAS-EXO-12-023} & \cite{Aad:2012em} & \cite{Khachatryan:2014hpa}  & \cite{Khachatryan:2014hpa}  & \none\\
$b$  &  & & & &  & \cite{CMS-PAS-EXO-12-023} & \cite{Aad:2015typ} & \none & \none  & \none\\
$t$ &  & & & & &  & \cite{Aad:2015fna} & \cite{Aad:2015voa} & \none  & \none\\
$W$ &  & & & & &  &  & \cite{Aad:2015agg,Aad:2015owa,Aad:2015ufa,Khachatryan:2014gha} & \cite{Aad:2015owa,Aad:2015ufa,Khachatryan:2014xja,Aad:2015ipg} & \cite{Aad:2015yza,Khachatryan:2016yji,Khachatryan:2015bma} \\
$Z$ &  & & & & &  & & & \cite{Aad:2015kna,Aad:2015owa,Khachatryan:2014gha} & \cite{Aad:2015yza,Khachatryan:2015lba,Khachatryan:2015ywa,Khachatryan:2015bma}  \\
$h$ &  & & & &  &  & & & & \cite{Aad:2015xja,Khachatryan:2015yea, CMS-PAS-EXO-15-008, Khachatryan:2016cfa} \\

\hline
\hline
\end{tabular}
\label{tab:res}
\end{table*}

\begin{table*}
\caption{ Theory models motivating two-body final state resonance searches. Here $Z'$ and $W'$ denote additional gauge bosons, $\not \! \! R$ denotes R-parity violating decays of sparticles in supersymmetry, $H^{\pm \pm}$ denotes doubly-charged Higgs bosons, $H$ denotes additional neutral scalar or pseudoscalar Higgs bosons, $L^*$ and $Q^*$ denote excited fermions, $X_{KK}$ denote various Kaluza-Klein excitations of gravitons or Standard Model fields, $\rho$ denotes neutral or charged techni-rhos, $LQ$ denotes leptoquarks, $T'$, $B'$, $Q'$ denote vector-like top, bottom, and light-flavor quarks, and $\mathcal{Q}$ denotes quirks. See also \cite{KCKong}. }
\begin{tabular}{lccccccccccc}
\hline
\hline
 & $e$\ \   & $\mu$\ \   & \ \  $\tau$\ \   & \ \  $\gamma$\ \   & \ \  $j$ \ \   & \ \  $b$\ \  & \ \ $t$ \ \ & \ \ $W$ \ \ & \ \ $Z$ \ \ & \ \ $h$ \ \   \\
\hline
$e$ &  $Z',H^{\pm\pm} $ & $\not \! \! R,H^{\pm\pm} $ & $\not \! \! R,H^{\pm\pm} $ & $L^*$  & $LQ,\not \!\! R$ & $LQ, \not \!\! R$ & $LQ,\not \!\! R$ & $L^*, \nu_{KK}$ & $L^*, e_{KK}$ & $L^*$ \\
$ \mu $ &  & $Z',H^{\pm\pm}$ & $\not \! \! R,H^{\pm\pm} $ & $L^*$  & $LQ,\not \!\! R$ & $LQ, \not \!\! R$ & $LQ,\not \!\! R$ & $L^*, \nu_{KK}$ & $L^*, \mu_{KK}$ & $L^*$ \\
$\tau$ &  &  & $Z',H,H^{\pm\pm}$ & $L^*$ & $LQ, \not \!\! R$ & $LQ,\not \!\! R$ & $LQ,\not \!\! R$ & $L^*, \nu_{KK}$ & $L^*, \tau_{KK}$ & $L^*$ \\
$\gamma$ &  & & & $H, G_{KK}, \mathcal{Q}$ & $Q^*$ & $Q^*$ & $Q^*$ & $W_{KK}, \mathcal{Q}$ & $H, \mathcal{Q}$ & $Z_{KK}$\\
$j$ &  &  &  &  & $Z',\rho, G_{KK}$ & $W',\not \!\! R$ & $T', \not \!\! R$ & $Q^*, Q_{KK}$  & $Q^*, Q_{KK}$  & $Q'$\\
$b$  &  &  &  &  &  & $Z',H$ & $W', \not \!\! R, H^\pm$ &  $T',Q^*, Q_{KK}$  & $Q^*, Q_{KK}$  & $B'$\\
$t$ &  &  &  &  &  &   & $H,G',Z'$ & $T'$ & $T'$  & $T'$\\
$W$ &  &  &  &  &  &  &  & $H, G_{KK}, \rho$ & $W', \mathcal{Q}$ & $H^\pm, \mathcal{Q}, \rho$ \\
$Z$ &  &  &  &  &  &  &  &  & $H, G_{KK}, \rho$ & $A, \rho$   \\
$h$ &  &  &  &  &  &  &  &  &  & $H, G_{KK}$ \\

\hline
\hline
\end{tabular}
\label{tab:restheory}
\end{table*}

\begin{table*}
\caption{The possible QCD and EM quantum numbers of each 2-body resonance, indicated as ({\bf QCD},{\bf EM}). Alternate quantum number assignments are indicated in parentheses. Round (square) brackets indicate a bosonic (fermionic) resonance. An ${}^*$ indicates that there is no possible initial state for resonant production at the LHC. A $\diamondsuit$ ($\heartsuit$) indicates that this state would lead to $\Delta B=1$ ($\Delta L=1$) processes if it possessed a resonant production mode at the LHC from additional couplings to quarks or gluons. \label{tab:quantumnumbers}}
\begin{center}
\begin{tabular}{cccccccccc} \hline\hline
& $\ell$ & $\gamma$ & $q$ & $g$ & $b$ & $t$ & $W^+$ & $Z$ & $h$ \\ \hline
$\ell$ & $\bnum{1}{2}^*$ & $\fnum{1}{1}^*$ & $\bnum{\bar 3}{\nicefrac{1(4)}{3}}^{\diamondsuit\heartsuit}$& $\fnum{8}{1}^*$& $\bnum{\bar 3}{\nicefrac{4}{3}}^{\diamondsuit\heartsuit}$& $\bnum{\bar 3}{\nicefrac{1}{3}}^{\diamondsuit\heartsuit}$& $\fnum{1}{0}^*$& $\fnum{1}{1}^*$ &$\fnum{1}{1}^*$  \\  
$\bar \ell$ & $\bnum{1}{0}$ & $\fnum{1}{-1}^*$ & $\bnum{\bar 3}{-\nicefrac{2(5^*)}{3}}^{\diamondsuit\heartsuit}$& $\fnum{8}{-1}^*$& $\bnum{\bar 3}{-\nicefrac{2}{3}}^{\diamondsuit\heartsuit}$& $\bnum{\bar 3}{-\nicefrac{5}{3}}^*$& $\fnum{1}{-2}^*$ & $\fnum{1}{-1}^*$& $\fnum{1}{-1}^*$   \\ 
$\gamma$ & $\fnum{1}{1}^*$ & $\bnum{1}{0}$ & $\fnum{\bar 3}{\nicefrac{1(-2)}{3}}$ & $\bnum{8}{0}$& $\fnum{\bar 3}{\nicefrac{1}{3}}$& $\fnum{\bar 3}{-\nicefrac{2}{3}}$ & $\bnum{1}{-1}$ & $\bnum{1}{0}$ &  $\bnum{1}{0}$ \\ 
$ q$ & $\bnum{\bar 3}{\nicefrac{1(4)}{3}}^{\diamondsuit\heartsuit}$ & $\fnum{\bar 3}{\nicefrac{1(-2)}{3}}$ & $\bnum{3}{\nicefrac{-1(2)(-4)}{3}}$ & $\fnum{\bar 3}{\nicefrac{1(-2)}{3}}$ & $\bnum{3}{\nicefrac{-1(2)}{3}}$ & $\bnum{3}{\nicefrac{-1(-4)}{3}}$ &  $\fnum{\bar 3}{\nicefrac{-2(-5^{*})}{3}}$&  $\fnum{\bar 3}{\nicefrac{1(-2)}{3}}$ &  $\fnum{\bar 3}{\nicefrac{1(-2)}{3}}$ \\ 
$ \bar q $ & $\bnum{3}{\nicefrac{2(5^*)}{3}}^{\diamondsuit\heartsuit}$ & $\fnum{3}{\nicefrac{-1(2)}{3}}$& $\bnum{1(8)}{0(-1)}$ & $\fnum{3}{\nicefrac{-1(2)}{3}}$ & $\bnum{1(8)}{0(-1)}$ & $\bnum{1(8)}{0(-1)}$  & $\fnum{3}{\nicefrac{-1(-4^*)}{3}}$ & $\fnum{3}{\nicefrac{-1(2)}{3}}$ & $\fnum{3}{\nicefrac{-1(2)}{3}}$  \\ 
$g$ & $\fnum{8}{1}^*$ & $\bnum{8}{0}$ & $\fnum{\bar 3}{\nicefrac{1(-2)}{3}}$ & $\bnum{1(8)}{0}$ & $\fnum{\bar 3}{\nicefrac{1}{3}}$ & $\fnum{\bar 3}{-\nicefrac{2}{3}}$ & $\bnum{8}{-1}$ &  $\bnum{8}{0}$ &  $\bnum{8}{0}$ \\ 
$b$ &  & $\fnum{\bar 3}{\nicefrac{1}{3}}$ & $\bnum{3}{\nicefrac{-1(2)}{3}}$ & $\fnum{\bar 3}{\nicefrac{1}{3}}$ & $\bnum{3}{\nicefrac{2}{3}}$ & $\bnum{3}{-\nicefrac{1}{3}}$ & $\fnum{\bar 3}{-\nicefrac{2}{3}}$ & $\fnum{\bar 3}{\nicefrac{1}{3}}$ & $\fnum{\bar 3}{\nicefrac{1}{3}}$   \\  
$\bar b$ & & & $\bnum{1(8)}{0(-1)}$ & $\fnum{3}{-\nicefrac{1}{3}}$ & $\bnum{1(8)}{0}$ & $\bnum{1(8)}{-1}$  & $\fnum{3}{-\nicefrac{4}{3}}^*$ & $\fnum{3}{-\nicefrac{1}{3}}$&  $\fnum{3}{-\nicefrac{1}{3}}$\\ 
$t$ & & &  & $\fnum{\bar 3}{-\nicefrac{2}{3}}$ & $\bnum{3}{-\nicefrac{1}{3}}$ & $\bnum{3}{-\nicefrac{4}{3}}$& $\fnum{\bar 3}{-\nicefrac{5}{3}}^*$ & $\fnum{\bar 3}{-\nicefrac{2}{3}}$ & $\fnum{\bar 3}{-\nicefrac{2}{3}}$  \\  
$\bar t$ & & & & & $\bnum{1(8)}{1}$ & $\bnum{1(8)}{0}$ & $\fnum{3}{-\nicefrac{1}{3}}$ & $\fnum{3}{\nicefrac{2}{3}}$& $\fnum{3}{\nicefrac{2}{3}}$  \\ 
$W^+$ & & &  & & & $\fnum{\bar 3}{-\nicefrac{5}{3}}^*$ & $\bnum{1}{-2}^*$ & $\bnum{1}{-1}$ & $\bnum{1}{-1}$ \\ 
$W^-$ & & & & & & & $\bnum{1}{0}$& $\bnum{1}{1}$& $\bnum{1}{1}$  \\ 
$Z$ & & & & & & & & $\bnum{1}{0}$ & $\bnum{1}{0}$  \\  
$h$ & & & & & & & & & $\bnum{1}{0}$  \\  \hline\hline

\end{tabular}
\end{center}
\label{default}
\end{table*}

\begin{table*}
\caption{For each pair of Standard Model particles, three boxes indicate the existence of various possible production modes for the corresponding resonance. In the first box, a \protect\resone \, indicates the existence of a resonant production mode at the LHC via the tree-level decay couplings, loop-induced processes involving the decay coupling, or the inclusion of additional couplings to quarks or gluons allowed by the quantum numbers of the resonance.  In the second box, \protect\restwo \,, \protect\resthree \,, \protect\resfour \,, or \protect\resfive \, indicate the leading production mode in association with one, two, three, or four Standard Model particles using the same coupling for production and decay in a four-flavor scheme. In the third box, \protect\resp \, indicates the {\it unavoidable} existence of a pair production mode via Standard Model gauge bosons. This box is left empty if there is a possible choice of resonance quantum numbers that does not lead to a pair production mode. }
\begin{center}
\def\arraystretch{1.5}
\begin{tabular}{c||c|c|c||c|c|c||c|c|c||c|c|c||c|c|c||c|c|c||c|c|c||c|c|c||c|c|c|} \hline

& \multicolumn{3}{|c||}{$\ell$} & \multicolumn{3}{|c||}{$\gamma$} & \multicolumn{3}{|c||}{$q$} & \multicolumn{3}{|c||}{$g$} & \multicolumn{3}{|c||}{$b$} & \multicolumn{3}{|c||}{$t$} & \multicolumn{3}{|c||}{$W^+$} & \multicolumn{3}{|c||}{$Z$} & \multicolumn{3}{|c|}{$h$} \\ \hline

$\ell$ &  & \resfive & \resp &  & \resfour  & \resp & \resone & \resthreea & \resp & & \resthreea & \resp & \resone & \resfour  & \resp &  \resone  &  \resfour  & \resp &  & \resfour  & &  & \resfour  & \resp &   & \resfour & \resp \\ \hline

$\bar \ell$ &  \resone & \resfive &  &  & \resfour  & \resp & \resone & \resthreea & \resp & & \resthreea & \resp & \resone & \resfour  & \resp & &  \resfour  & \resp  &  & \resfour  & \resp &  & \resfour  & \resp &  & \resfour  & \resp \\ \hline

$\gamma$ &  & \resfour & \resp & \resone & \resthree &  & \resone & \restwo & \resp & \resone & \restwo & \resp & \resone  & \resthree & \resp &  \resone  & \resthree & \resp   & \resone & \resthree & \resp & \resone & \resthree &  & \resone & \resthree &  \\ \hline 

$ q $ &  \resone  &  \resthree  &  \resp   & \resone & \restwo & \resp &  \resone  &  \resone  &  \resp  &  \resone  &  \resone  &  \resp   &  \resone  &  \restwo  & \resp &  \resone  &  \restwo  & \resp  & \resone & \restwo & \resp & \resone & \restwo & \resp & \resone & \restwo & \resp   \\ \hline

$\bar q$ &  \resone  &  \resthree  &  \resp  & \resone & \restwo & \resp &  \resone  &  \resone  &  \resp  &  \resone  &  \resone  &  \resp   &   \resone  &  \restwo  &  &  \resone  &  \restwo  &   &  \resone  &  \restwo  &  \resp  & \resone & \restwo & \resp & \resone & \restwo & \resp  \\ \hline

$g$ &    &  \resthree  & \resp  & \resone & \restwo & \resp &  \resone  &  \resone  & \resp  &  \resone  &  \resone  &  &  \resone  &  \restwo  & \resp  &  \resone  &  \restwo  & \resp  &   \resone  &  \restwo  &  \resp  &   \resone  &  \restwo  &  \resp  & \resone & \restwo & \resp \\ \hline

$b$ &  && & \resone  & \resthree & \resp &  \resone  &  \restwo  & \resp  & \resone  &  \restwo  & \resp &  \resone  &  \resthree  & \resp  &  \resone  &  \resthree  & \resp  & \resone & \resthree & \resp & \resone  & \resthree & \resp & \resone  & \resthree & \resp \\ \hline  

$ \bar b$ & &&&&&  &  \resone  &  \restwo  &   &  \resone  &  \restwo  & \resp &  \resone  &  \resthree  &  &  \resone  &  \resthree  & \resp  &   & \resthree & \resp &  \resone  & \resthree & \resp  &  \resone  & \resthree & \resp   \\ \hline

$t$ & &&&&&&&& & \resone  &  \restwo  & \resp &  \resone & \resthree & \resp & \resone & \resthree & \resp & &  \resthree  &  \resp & \resone  &  \resthree  & \resp & \resone   &  \resthree  & \resp \\ \hline

$\bar t$ & &&&&&&&&&&& &  \resone  &   \resthree  & \resp  &  \resone  &  \resthree  &     &  \resone  &  \resthree  & \resp & \resone  &  \resthree  & \resp & \resone   &  \resthree  & \resp \\ \hline

$W^+ $ & &&&&&&&&&&&&&& & &  \resthree  &  \resp &  & \resthree & \resp & \resone & \resthree & \resp & \resone & \resthree & \resp \\ \hline

$W^-$ &  &&&&&&&&&&&&&&&&& & \resone & \resthree &  & \resone & \resthree & \resp & \resone & \resthree & \resp   \\ \hline

$Z$ &  &&&&&&&&&&&&&&&&&&&& & \resone & \resthree &  & \resone & \resthree &   \\ \hline

$h$ & &&&&&&&&&&&&&&&&&&&&&&& & \resone & \resthree &  \\ \hline

\end{tabular}
\end{center}
\label{tab:modes}
\end{table*}%

\section{Discussion}

The data from the LHC are extraordinarily valuable, in that its collection required an enormous investment of financial and human resources and in its potential power to answer outstanding questions of particle physics.  However, once those resources are spent and the data are collected, there remain difficult questions regarding how to use it. Experimental analysis of a given final state requires limited human and financial resources, and every search increases field-wide trials factor, making any local excess less globally significant.  Therefore, it is necessarily the case that some experimental territory will be left uncovered, and proposals for new experimental searches must have a  compelling argument.

Here we have argued that in addition to the usual stable of theoretically-motivated searches, a set of experimentally-motivated searches should be conducted.  We propose a set of exclusive 2-body resonance searches, which naturally limits the number of final states and are well matched to experimental capabilities. This is in contrast to the strategy of general searches, which attempt to satisfy a broad set of theory motivations, but do not focus on experimental strengths and suffer a very large trials factor.

The final states with matched objects have been examined, though there remain openings at low- and high-mass regions. More significantly, we find that many of the mismatched pair final states have had no attention, despite the existence of theoretical models and the absence of strong theoretical constraints. \\

\section*{Acknowledgements}

We thank Mohammad Abdullah, Jahred Adelman, and Tim Tait for useful conversations. This research was supported in part by the National Science Foundation under Grant No. NSF PHY11-25915 and the US Department of Energy under the grant DE-SC0014129 and DE-FG02-12ER41809; the authors are grateful to the Kavli Institute For Theoretical Physics, where some of this work was done.

\bibliography{topo}

\end{document}